\documentclass[10pt, pre, aps, twocolumn,showpacs,nofootinbib,longbibliography]{revtex4-1}
\usepackage{url,ulem}
\usepackage{amsmath,amssymb}
\usepackage[dvipdfmx]{graphicx,color}

\newcommand{\mX}{\boldsymbol{\rm X}}
\newcommand{\mK}{\boldsymbol{\rm K}}
\newcommand{\mA}{\boldsymbol{\rm A}}
\newcommand{\mB}{\boldsymbol{\rm B}}
\newcommand{\mC}{\boldsymbol{\rm C}}
\newcommand{\mP}{\boldsymbol{\rm P}}

\newcommand{\mLambda}{\boldsymbol{\rm \Lambda}}
\newcommand{\mW}{\boldsymbol{\rm W}}

\newcommand{\Vspring}{V_{\rm spring}}
\newcommand{\Vbend}{V_{\rm bend}}
\newcommand{\Leff}{L_{{\rm eff}}}
\newcommand{\Meff}{M_{{\rm eff}}}
\newcommand{\Veff}{V_{{\rm eff}}}

\newcommand{\br}{\boldsymbol{r}}
\newcommand{\bq}{\boldsymbol{q}}
\newcommand{\bp}{\boldsymbol{p}}
\newcommand{\be}{\boldsymbol{e}}
\newcommand{\bx}{\boldsymbol{x}}

\newcommand{\by}{\boldsymbol{y}}
\newcommand{\bz}{\boldsymbol{z}}
\newcommand{\bzero}{\boldsymbol{0}}
\newcommand{\bbeta}{\boldsymbol{\eta}}

\newcommand{\norm}[1]{||#1||}
\newcommand{\dfracd}[2]{\dfrac{{\rm d} #1}{{\rm d} #2}}
\newcommand{\dfracp}[2]{\dfrac{\partial #1}{\partial #2}}
\newcommand{\dfracpp}[3]{\dfrac{\partial^{2} #1}{\partial #2\partial #3}}
\newcommand{\ave}[1]{\left\langle #1 \right\rangle}
\newcommand{\Tr}{{\rm Tr}}

\newcommand{\fullword}{\textcolor{black}{dynamically induced conformation}}
\newcommand{\word}{\textcolor{black}{DIC}}

\begin{document}

\title{Dynamically induced conformation depending on excited normal modes of fast oscillation}
\author{Yoshiyuki Y. Yamaguchi}
\email{yyama@amp.i.kyoto-u.ac.jp}
\affiliation{Department of Applied Mathematics and Physics, Graduate School of Informatics, Kyoto University, Kyoto 606-8501, Japan}
\author{Tatsuo Yanagita}
\affiliation{Department of Engineering Science, Osaka Electro-Communication University, Neyagawa 572-8530, Japan}
\author{Tetsuro Konishi}
\affiliation{General Education Division, College of Engineering, Chubu University, Kasugai 487-8501, Japan}
\author{Mikito Toda}
\affiliation{Faculty Division of Natural Sciences, Research Group of Physics, Nara Women's University, Kita-Uoya-Nishimachi, Nara 630-8506, Japan}
\email{Present position:  Research fellow, Nara Women's University}
\affiliation{Graduate School of Information Science , University of Hyogo, 7-1-28 Minatojima-minamimachi, Chuo-ku, Kobe, Hyogo 650-0047, Japan}
\affiliation{Research Institute for Electronic Science, Hokkaido University, Kita 20 Nishi 10, Kita-Ku, Sapporo 001-0020, Japan}

\begin{abstract}

  We present dynamical effects on conformation in a simple bead-spring model
  consisting of three beads connected by two stiff springs.
  The conformation defined by the bending angle between the two springs
  is determined not only by a given potential energy function
  depending on the bending angle,
  but also fast motion of the springs which constructs the effective potential.
  A conformation corresponding with a local minimum of the effective potential
  is hence called the dynamically induced conformation.
  We develop a theory to derive the effective potential
  by using multiple-scale analysis and the averaging method.
  A remarkable consequence is that the effective potential depends on
  the excited normal modes of the springs and amount of the spring energy.
  Efficiency of the obtained effective potential is numerically verified.
  
\end{abstract}
\maketitle

\section{Introduction}

Conformation is deeply connected with function.
A typical example is  a biomolecule
whose conformation is crucial for binding a ligand
\cite{koshland-58,monod-wyman-changeux-65,okazaki-takada-08,seeliger-degroot-10,fuchigami-etal-11}.
Morphological computation \cite{hauser-etal-11,special-issue-13,muller-hoffmann-17}
is another example, which can be found for instance as walking robots
\cite{collins-ruina-tedrake-wisse-05,hermans-schrauwen-bienstman-dambre-14}.
Mechanical metamaterial \cite{mechanical-metamaterials-19} provides
several examples like the Miura fold which exhibits negative Poisson's ratio
\cite{wei-etal-13}.

Realization of conformations is 
usually associated with the minimum of a potential energy function.
In addition to the potential function,
dynamics sometimes contributes
to construct an effective potential.
A well-known example is the Kapitza pendulum \cite{stephenson-08,kapitza-51}:
An inverted pendulum persists against the gravity
by applying a rapidly oscillating external force,
since the effective potential provides a local minimum at the inverted position.

We present another dynamical effect on conformation
realized in autonomous Hamiltonian systems
containing fast and slow motion.
Consider a bead-spring model \cite{rouse-53}
consisting of three beads connected by two stiff springs.
The conformation of this system can be identified
with the bending angle between the two springs.
If the system has a bending potential which depends only on the bending angle,
one may imagine that the bending angle goes to a local minimum
of the bending potential.
The conformation of this system is, however, determined
by the effective potential which consists of the bending potential
and contribution from the fast spring motion. 

More precise description of the above phenomenon is as follows.
First of all, the dynamical effect is comparable with the bending potential
under the condition that large bending motion is sufficiently slow than
the spring motion.
The bending potential dominates the effective potential
when the spring energy is sufficiently small.
However, the dynamical effect enlarges as the spring energy increases,
and a local minimum of the effective potential does not necessarily
coincide with a local minimum of the bending potential.
We call a conformation corresponding to a local minimum
of the effective potential a {\fullword} ({\word}).
Further interesting fact is that the effective potential depends
on the excited normal modes of the springs in addition to the spring energy.
The three-body bead-spring model has the two normal modes of the springs
and each mode makes a different valley.

The three-body system is quite simple,
and hence it is theoretically tractable and clearly shows DIC.
The aim of this paper is to present 
the dynamical contribution to conformation in the three-body system.
It is worth noting that the bead-spring model mimics several systems:
a polymer \cite{rouse-53,zimm-56}, 
a microscopic artificial swimmer \cite{gauger-stark-06},
a soft magnetic nanowire \cite{mirzae-etal-20},
and a semi-flexible macromolecule \cite{saadat-khomami-16}.

A theoretical analysis reveals that the essence of DIC
is existence of multiple timescales,
which is realized in the bead-spring model by stiff springs and slow bending motion.
Appearance of multiple timescales is generic in nature.
For instance, biomolecules have several forces of diverse strength
as strong covalent bonds, intermediate hydrogen bonds, and weak van der Waals forces,
and each of them has a characteristic timescale.
{\word} therefore enriches understanding of the origin of conformation change
and its function.

A celebrated example of the dynamically constructed effective potential
is found in the aforementioned Kapitza pendulum,
and it has been studied in a wide range of fields
\cite{bukov-dalessio-polkovnikov-15,grifoni-hanggi-96,wickenbrock-etal-12,chizhevsky-smeu-giacomelli-03,chizhevsky-14,cubero-etal-06,bordet-morfu-13,weinberg-14,uzuntarla-etal-15,buchanan-jameson-oedjoe-62,baird-63,jameson-66,apffel-20,bellman-mentsman-meerkov-86,shapiro-zinn-97,borromeo-marchesoni-07,richards-etal-18}.
The development of the Kapitza pendulum suggests
that DIC will provide a large spectrum of applications.
Nevertheless, we underline three crucial differences between
{\word} and the Kapitza pendulum.
(i) The bead-spring model is autonomous
and the effective potential is intrinsically determined,
while one in the Kapitza pendulum can be controlled
by the applied external force.
(ii) The effective potential depends on the excited normal modes
in the bead-spring model.
(iii) The local minimum points of the effective potential
may continuously move depending on the spring energy in the bead-spring model,
while the local minimum created by the external vertical oscillation
is fixed at the inverted position in the Kapitza pendulum.

This paper is organized as follows.
The three-body bead-spring model is presented in Sec.~\ref{sec:model}
with the two important assumptions to have DIC.
Following the assumptions, we develop in Sec.~\ref{sec:theory}
a theory to describe slow bending motion
by using a multiple-scale analysis \cite{bender-orszag-99}
and the averaging method
\cite{krylov-bogoliubov-34,krylov-bogoliubov-47,guckenheimer-holmes-83}.
The theory provides the effective potential
depending on the excited normal modes of the springs and amount of the spring energy.
Examples of the effective potential are exhibited in Sec.~\ref{sec:effective-potentials}
so as to reveal the above dependency.
Efficiency of the effective potential is examined through numerical simulations
in Sec.~\ref{sec:numerics}.
The final section \ref{sec:summary} is devoted to summary and discussions.

\section{Model}
\label{sec:model}

\begin{figure}
  \centering
  \includegraphics[width=5cm]{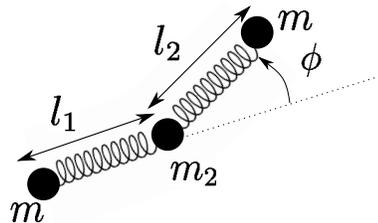}
  \caption{The three-body bead-spring model.
    We assume the symmetric masses $m_{1}=m_{3}=m$
    and the two springs having the idential potential.}
  \label{fig:Rouse3model}
\end{figure}

The three-body bead-spring model is sketched in Fig.~\ref{fig:Rouse3model}.
We assume that the beads move on a two-dimensional plane.
The mass and the position of the $j$th bead are respectively
denoted by $m_{j}$ and $\br_{j}\in\mathbb{R}^{2}$.
The Lagrangian of the model is expressed by
\begin{equation}
  \label{eq:model}
  L = \dfrac{1}{2} \sum_{j=1}^{3} m_{j} \norm{\dot{\br}_{j}}^{2}
  - V(\br_{1},\br_{2},\br_{3}),
\end{equation}
where $\dot{\br}_{j}:=\mathrm{d}\br_{j}/\mathrm{d}t$
and $\norm{\cdot}$ is the Euclidean norm:
$\norm{\br}=\sqrt{x^{2}+y^{2}}$ for $\br=(x,y)\in\mathbb{R}^{2}$.
The $j$th and the $(j+1)$th beads are connected by a stiff spring,
and we assume that the two springs have the identical potential.
Further, for simplicity, we focus on the symmetric masses: $m_{1}=m_{3}=m$.
The term $V$ represents the potential energy function, which will be specified later.

We assume that the system described by Eq.~\eqref{eq:model} has
the two-dimensional translational symmetry and the rotational symmetry,
which induce the conservation of the two-dimensional total momentum vector
and of the total angular momentum, respectively.
The total momentum vector can be set as the zero vector without loss of generality,
while the total angular momentum is assumed to be zero.
The three integrals reduces the system and the reduced Lagrangian is
\begin{equation}
  \label{eq:Lagrangian-y}
  L
  = \dfrac{1}{2} \sum_{\alpha,\beta=1}^{3}
  C^{\alpha\beta}(\by) \dot{y}_{\alpha} \dot{y}_{\beta}
  - V(\by),
\end{equation}
where
\begin{equation}
  \by = (y_{1},y_{2},y_{3})^{\rm T} = (l_{1},l_{2},\phi)^{\rm T}
\end{equation}
and the superscript T represents transposition.
The variables $l_{1}$ and $l_{2}$ are the lengths of the two springs,
\begin{equation}
  l_{j} = \norm{ \br_{j+1}-\br_{j} }, \quad (j=1,2) 
\end{equation}
and $\phi$ is the bending angle,
\begin{equation}
  \cos\phi = \dfrac{(\br_{3}-\br_{2})\cdot(\br_{2}-\br_{1})}{\norm{\br_{3}-\br_{2}}\norm{\br_{2}-\br_{1}}},
\end{equation}
where $\cdot$ is the Euclidean inner product.
The function $C^{\alpha\beta}(\by)$ is the $(\alpha,\beta)$ element
of the size-$3$ symmetric matrix $\mC(\by)$,
whose explicit form is given in Appendix \ref{sec:three-body-Rouse-model}.
The potential energy function $V(\by)$ consists of the two parts as
\begin{equation}
  \label{eq:Vspring-Vbend}
  V(\by) = \Vspring(l_{1},l_{2}) + \Vbend(\phi).
\end{equation}
We call $\Vspring$ and $\Vbend$ the spring potential and the bending potential,
respectively.

We introduce the two assumptions to realize {\word} in the above model.
Let $\epsilon$ be a dimensionless small parameter as $|\epsilon|\ll 1$.
The two assumptions are:
\begin{itemize}
\item[({\it A1})] The amplitudes of the springs are sufficiently small
  comparing with the natural length. The ratio is of $O(\epsilon)$.
\item[({\it A2})] Large bending motion is sufficiently slow than the spring motion.
  The ratio of the two timescales is of $O(\epsilon)$.
\end{itemize}
These two assumptions lead the effective potential for the bending angle $\phi$.
A local minimum point of the effective potential does not necessarily coincide with
a local minimum point of the bending potential $\Vbend(\phi)$.
That is, the bending angle oscillates around 
an angle at which the bending potential $\Vbend(\phi)$ does not take
a local minimum.
We develop a theory to derive the effective potential in Sec.\ref{sec:theory}.

\section{Theory}
\label{sec:theory}

From now on, we use the Einstein notation for the sum:
We take the sum over an index if it appears twice in a term.
We derive the equation of motion for the slow bending motion,
and construct the effective potential induced by the fast spring motion.
A review of the Kapitza pendulum is provided in Appendix
\ref{eq:Kapitza-pendulum},
which might be helpful to understand the theory.

\subsection{Multiscale analysis and averaging}

The Euler-Lagrange equations derived from Eq.~\eqref{eq:Lagrangian-y} are
\begin{equation}
  \label{eq:Euler-Lagrange}
  C^{\alpha\beta}(\by) \ddot{y}_{\beta}
  + \left[
    \dfracp{C^{\alpha\beta}}{y_{\gamma}}(\by)
    - \dfrac{1}{2} \dfracp{C^{\beta\gamma}}{y_{\alpha}}(\by)
  \right]
  \dot{y}_{\beta} \dot{y}_{\gamma}
  + \dfracp{V}{y_{\alpha}}(\by) = 0,
\end{equation}
where $\alpha,\beta,\gamma\in\{1,2,3\}$.
These equations are the starting point of our theory.

The assumption ({\it A2}) induces the two timescales of $t_{0}=t$ and $t_{1}=\epsilon t$.
The fast timescale $t_{0}$ describes the fast spring motion,
and the slow timescale $t_{1}$ corresponds to the slow bending motion.
The two timescales transform the time derivative into
\begin{equation}
  \label{eq:expansion-time}
  \dfracd{}{t} = \dfracp{}{t_{0}} + \epsilon \dfracp{}{t_{1}}.
\end{equation}

From the assumptions ({\it A1}) and ({\it A2}) the variables $l_{j}$ and $\phi$
are expanded as
\begin{equation}
  \label{eq:l-phi-expansion}
  \left\{
    \begin{split}
      & l_{j}(t_{0},t_{1}) = l_{j}^{(0)} + \epsilon l_{j}^{(1)}(t_{0},t_{1}),
      \quad l_{j}^{(0)}=l_{\ast}
      \quad (j=1,2) \\
      & \phi(t_{0},t_{1}) = \phi^{(0)}(t_{1}) + \epsilon \phi^{(1)}(t_{0},t_{1}),
    \end{split}
  \right.
\end{equation}
where $l_{\ast}$ is the natural length of the two springs.
As we will observe later, the fast motion of $\phi^{(1)}(t_{0},t_{1})$ is induced
by the fast motion of the springs and is of the same order $O(\epsilon)$
as the spring amplitudes.
We are interested in $\phi^{(0)}(t_{1})$,
which represents large and slow bending motion.
We denote the above expansions for simplicity as
\begin{equation}
  \label{eq:expansion-space}
  \by(t_{0},t_{1}) = \by^{(0)}(t_{1}) + \epsilon \by^{(1)}(t_{0},t_{1}).
\end{equation}

We further expand the potential energy function $V$.
The spring potential $\Vspring$ is assumed to be expanded into the Taylor series
around the natural length as
\begin{equation}
  \label{eq:expansion-Vspring}
  \Vspring(l_{1},l_{2})
  = \dfrac{k}{2} \sum_{j=1}^{2} \left( l_{j} - l_{\ast} \right)^{2}
  + O(|l_{j}-l_{\ast}|^{3}).
\end{equation}
That is, the two springs have the same spring constant
\begin{equation}
  k = \dfracp{{}^{2}\Vspring}{l_{j}^{2}}(l_{\ast},l_{\ast}) \quad (j=1,2).
\end{equation}
The bending potential is assumed to be expanded into the series of $\epsilon$ as
\begin{equation}
  \label{eq:expansion-Vbend}
  \Vbend(\phi)
  = \Vbend^{(0)}(\phi)
  + \epsilon \Vbend^{(1)}(\phi)
  + \epsilon^{2} \Vbend^{(2)}(\phi)
  + \cdots.
\end{equation}
The two assumptions ({\it A1}) and ({\it A2}) induce
\begin{equation}
  \Vbend^{(0)}(\phi), ~ \Vbend^{(1)}(\phi)\equiv 0
\end{equation}
as shown in Appendix \ref{sec:Vbend-Oepsilon2},
and hence the leading term of $\Vbend$ is of $O(\epsilon^{2})$.
This ording results from the assumption {\it (A2)}:
The force from the bending potential $\Vbend$
should be weaker than that of the spring potential $\Vspring$.

We construct the equations of motion order by order,
by substituting Eqs.~\eqref{eq:expansion-time}, \eqref{eq:expansion-space}, 
\eqref{eq:expansion-Vspring}, and \eqref{eq:expansion-Vbend}
into Eq.~\eqref{eq:Euler-Lagrange}.
In $O(\epsilon^{0})$, we have no terms,
because $\partial y_{\beta}^{(0)}/\partial t_{0}=0$,
$\dot{y}_{\beta}, \ddot{y}_{\beta}=O(\epsilon)$,
and $\partial V/\partial y_{\alpha}=O(\epsilon)$.

In $O(\epsilon)$ we have
\begin{equation}
  \label{eq:Oepsilon1}
  \dfracp{{}^{2}\by^{(1)}}{t_{0}^{2}} = - \mX(\by^{(0)}) \by^{(1)}.
\end{equation}
The size-$3$ matrix $\mX$ is defined by
\begin{equation}
  \mX(\by) = [\mC(\by)]^{-1} \mK,
\end{equation}
where
\begin{equation}
  \mK =
  \begin{pmatrix}
    k & 0 & 0 \\
    0 & k & 0 \\
    0 & 0 & 0 \\
  \end{pmatrix}.
\end{equation}
See Appendix \ref{sec:matrices-Oepsilon1} for the explicit form of $\mX(\by^{(0)})$.
The third column vector of $\mX$ is the zero vector,
and the right-hand side of Eq.~\eqref{eq:Oepsilon1} has no contribution from
the third element of $\by^{(1)}$, namely $\phi^{(1)}$.
The fast motion of $\phi^{(1)}$ is hence induced by $l_{1}^{(1)}$ and $l_{2}^{(1)}$,
as mentioned after Eq.~\eqref{eq:l-phi-expansion}.

The slow motion of $\phi^{(0)}(t_{1})$ is described in $O(\epsilon^{2})$,
and the equation of motion for $\phi^{(0)}(t_{1})$ is
\begin{equation}
  \label{eq:Oepsilon2-phi}
  \begin{split}
    & C^{3\beta}(\by^{(0)}) (\ddot{y}_{\beta})^{(2)}
    + \left[
      \dfracp{C^{3\beta}}{y_{\gamma}}(\by^{(0)})
      - \dfrac{1}{2} \dfracp{C^{\beta\gamma}}{\phi}(\by^{(0)})
    \right] (\dot{y}_{\beta})^{(1)} (\dot{y}_{\gamma})^{(1)} \\
    & + \dfracp{C^{3\beta}}{y_{\gamma}}(\by^{(0)})
    ( \ddot{y}_{\beta} )^{(1)} y_{\gamma}^{(1)}
    + \dfracd{\Vbend^{(2)}}{\phi}(\phi^{(0)}) = 0.
  \end{split}
\end{equation}
Here $(\dot{\by})^{(1)}$ is the first order part of $\dot{\by}$
and $(\dot{\by})^{(1)}\neq {\rm d}\by^{(1)}/{\rm d}t$.
The explicit forms are
\begin{equation}
  \begin{split}
    & (\dot{\by})^{(1)} = \dfracd{\by^{(0)}}{t_{1}} + \dfracp{\by^{(1)}}{t_{0}}, \\
    & (\ddot{\by})^{(1)} = \dfracp{{}^{2}\by^{(1)}}{t_{0}^{2}}, \\
    & (\ddot{\by})^{(2)} = \dfracd{{}^{2}\by^{(0)}}{t_{1}^{2}} + 2 \dfracpp{\by^{(1)}}{t_{0}}{t_{1}}. \\
  \end{split}
\end{equation}

Equation \eqref{eq:Oepsilon2-phi} contains the fast oscillation of $\by^{(1)}$,
and we eliminate it by taking the average over the fast timescale $t_{0}$.
The average is defined by
\begin{equation}
  \ave{\varphi}(t_{1}) = \lim_{T\to\infty} \dfrac{1}{T} \int_{0}^{T} \varphi(t_{0},t_{1}) dt_{0}
\end{equation}
for an arbitrary function $\varphi(t_{0},t_{1})$.
After taking the average and recalling $\by^{(0)}=(l_{\ast},l_{\ast},\phi^{(0)})$,
Eq.~\eqref{eq:Oepsilon2-phi} is simplified to
\begin{equation}
  \label{eq:Oepsilon2-averaged}
  \begin{split}
    & C^{33}(\by^{(0)}) \dfracd{{}^{2}\phi^{(0)}}{t_{1}^{2}}
    + \dfrac{1}{2} \dfracp{C^{33}}{\phi}(\by^{(0)})
    \left( \dfracd{\phi^{(0)}}{t_{1}} \right)^{2} \\
    & = - \dfracd{\Vbend^{(2)}}{\phi}(\phi^{(0)})
    + \mathcal{A}.
  \end{split}
\end{equation}
The right-hand side represents the force, and the averaged term
\begin{equation}
  \label{eq:average-A-0}
  \mathcal{A} = \dfrac{1}{2} \Tr \left[
    \dfracp{\mC}{\phi}(\by^{(0)})
    \ave{ \dfracp{\by^{(1)}}{t_{0}} \left( \dfracp{\by^{(1)}}{t_{0}} \right)^{\rm T} }
  \right]
\end{equation}
represents the effective force yielded by the fast spring motion.
Here Tr represents the matrix trace.
The right-hand side of Eq.~\eqref{eq:average-A-0} depends on
$\by^{(0)}$ and $\by^{(1)}$.
The $\by^{(0)}$ dependence can be regarded as the $\phi^{(0)}$ dependence,
since $\by^{(0)}=(l_{\ast},l_{\ast},\phi^{(0)})$ and $l_{\ast}$ is constant.  
$\by^{(1)}$ depends on $t_{0}$ and $t_{1}$,
and the $t_{0}$ dependence is averaged out.
We have to eliminate the $t_{1}$ dependence to obtain the effective potential
as a function of $\phi^{(0)}$.

\subsection{Explicit form of the averaged term}
\label{sec:explicit-form-A}

We compute the explicit form of the averaged term $\mathcal{A}$
by performing the diagonalization of Eq.~\eqref{eq:Oepsilon1},
and observe the $t_{1}$ dependence, which has to be eliminated.
Let $\mP$ diagonalize $\mX$ as
\begin{equation}
  \mX(\by^{(0)}) \mP(\by^{(0)}) = \mP(\by^{(0)}) \mLambda(\by^{(0)}).
\end{equation}
The diagonal matrix $\mLambda(\by^{(0)})$
consists of the eigenvalues of $\mX(\by^{(0)})$
and is denoted by
\begin{equation}
  \mLambda(\by^{(0)}) =
  \begin{pmatrix}
    \lambda_{1}(\by^{(0)}) & 0 & 0 \\
    0 & \lambda_{2}(\by^{(0)}) & 0 \\  
    0 & 0 & 0 \\
  \end{pmatrix},
\end{equation}
where
\begin{equation}
  \label{eq:X-eigenvalues}
  \begin{split}
    & \lambda_{1}(\by^{(0)}) = \dfrac{k(M_{2}-M_{1}\cos\phi^{(0)})}{M_{2}^{2}-M_{1}^{2}}, \\
    & \lambda_{2}(\by^{(0)}) = \dfrac{k(M_{2}+M_{1}\cos\phi^{(0)})}{M_{2}^{2}-M_{1}^{2}}, \\
  \end{split}
\end{equation}
and
\begin{equation}
  \label{eq:M2M1}
  M_{2} = \dfrac{m(m+m_{2})}{2m+m_{2}},
  \quad
  M_{1} = \dfrac{m^{2}}{2m+m_{2}}.
\end{equation}
A diagonalizing matrix is
\begin{equation}
  \label{eq:P}
  \mP(\by^{(0)}) =
  \begin{pmatrix}
    \bp_{\rm in}, & \bp_{\rm anti}, & \bp_{\phi} \\
  \end{pmatrix}
  = 
  \begin{pmatrix}
    1/\sqrt{2} & 1/\sqrt{2} & 0 \\
    1/\sqrt{2} & -1/\sqrt{2} & 0 \\
    v(\by^{(0)}) & 0 & 1 \\
  \end{pmatrix}
\end{equation}
with
\begin{equation}
  v(\by^{(0)})
  = \dfrac{\sqrt{2}}{l_{\ast}} \dfrac{M_{1}\sin\phi^{(0)}}{M_{2}-M_{1}\cos\phi^{(0)}}.
\end{equation}
The three column vectors $\bp_{\rm in}, \bp_{\rm anti}$, and $\bp_{\phi}$
are eigenvectors of $\mX(\by^{(0)})$,
and we call the three modes as the in-phase mode, the anti-phase mode,
and the zero-eigenvalue mode, respectively.

To solve Eq.~\eqref{eq:Oepsilon1},
we perform the change of variables as
\begin{equation}
  \by^{(1)} = \mP(\by^{(0)}) \bbeta,
\end{equation}
and $\bbeta$ solves the diagonalized equations
\begin{equation}
  \dfracp{{}^{2}\bbeta}{t_{0}^{2}} = - \mLambda(\by^{(0)}) \bbeta.
\end{equation}
Denoting the amplitudes of the in-phase and the anti-phase modes
by $w_{1}(t_{1})$ and $w_{2}(t_{1})$ respectively,
which evolve in the slow timescale $t_{1}$
through the coupling with $\phi^{(0)}(t_{1})$,
and setting the amplitude of the zero-eigenvalue mode as zero,
we introduce the diagonal matrix
\begin{equation}
  \mW(t_{1}) =
  \begin{pmatrix}
    w_{1}(t_{1}) & 0 & 0 \\
    0 & w_{2}(t_{1}) & 0 \\
    0 & 0 & 0 \\
  \end{pmatrix}.
\end{equation}

Putting all together and remembering that the average
of the square of a sinusoidal function is $1/2$, we have
\begin{equation}
  \label{eq:averageA}
  \begin{split}
    & \mathcal{A}(\phi^{(0)},w_{1},w_{2})
    = \dfrac{1}{4} \Tr \left[
      \dfracp{\mC}{\phi}(\by^{(0)}) \mP \mLambda \mW^{2} \mP^{\rm T} \right] \\
    & = - \dfrac{k}{4} \left[
      \dfrac{M_{1}\sin\phi^{(0)}}{M_{2}-M_{1}\cos\phi^{(0)}} w_{1}^{2}
      - \dfrac{M_{1}\sin\phi^{(0)}}{M_{2}+M_{1}\cos\phi^{(0)}} w_{2}^{2}
    \right].
  \end{split}
\end{equation}
The averaged term $\mathcal{A}$ contains
the two evolving amplitudes $w_{1}(t_{1})$ and $w_{2}(t_{1})$.
The untrivial evolution of the amplitudes differs from the Kapitza pendulum,
which also contains the amplitude of the external oscillation
but it is explicitly given.
We have to eliminate the two unknown amplitudes from the averaged term $\mathcal{A}$
to obtain a closed equation for $\phi^{(0)}(t_{1})$.

\subsection{Hypothesis and energy conservation}

The strategy to eliminate the two unknown amplitudes $w_{1}(t_{1})$ and $w_{2}(t_{1})$
is as follows.
First, we introduce a hypothesis, which is inspired from the adiabatic invariant.
The hypothesis reduces the number of unknown variables from two to one.
Second, we eliminate the remaining unknown variable
by using the energy conservation law.

The first step is the introduction of the hypothesis expressed by
\begin{equation}
  \label{eq:hypothesis}
  ({\it H})
  \quad
  w_{1}(t_{1})^{2} = \nu_{1} w(t_{1})^{2},
  \quad
  w_{2}(t_{1})^{2} = \nu_{2} w(t_{1})^{2},
\end{equation}
where $\nu_{1}$ and $\nu_{2}$ are constants satisfying
\begin{equation}
  \label{eq:nu1-nu2}
  \nu_{1} + \nu_{2} = 1.
\end{equation}
A physical interpretation of the hypothesis ({\it H}) is
that the slow bending motion exchanges energy with the fast spring motion
in proportion to its normal mode energy.
Validity of the hypothesis ({\it H}) is examined
in Appendix \ref{sec:hypothesis-check}.
We note that the hypothesis should be valid
if the modification of $\phi^{(0)}$ is sufficiently small.
The unique unknown variable is now $w(t_{1})$,
while the constants $\nu_{1}$ and $\nu_{2} $ have been included
in the equations of motion.

The second step is the energy conservation.
The leading order of the total energy is of $O(\epsilon^{2})$
and we expand it as $E=\epsilon^{2}E^{(2)}+O(\epsilon^{3})$.
The leading term is
\begin{equation}
  \begin{split}
    E^{(2)}
    & = \dfrac{1}{2} \Tr \left[ \mC(\by^{(0)}) (\dot{\by})^{(1)} (\dot{\by})^{(1){\rm T}} \right] \\
    & + \dfrac{1}{2} \Tr \left[ \mK \by^{(1)} \by^{(1){\rm T}} \right]
    + \Vbend^{(2)}(\phi^{(0)}).
  \end{split}
\end{equation}
Taking the average over $t_{0}$, we have
\begin{equation}
  \begin{split}
    \ave{E^{(2)}}
    & = \dfrac{1}{2} \Tr \left[ \mC(\by^{(0)}) \dfracd{\by^{(0)}}{t_{1}}
      \left( \dfracd{\by^{(0)}}{t_{1}} \right)^{\rm T} \right] \\
    & + \dfrac{k}{2} \Tr \mW^{2} 
    + \Vbend^{(2)}(\phi^{(0)}).
  \end{split}
\end{equation}
Substituting Eq.~\eqref{eq:hypothesis}, the unique unknown variable $w$
is obtained as
\begin{equation}
  \label{eq:w}
  k w(t_{1})^{2}
  = 2 \left[ E^{(2)} - \Vbend^{(2)}(\phi^{(0)}) \right]
  - C^{33}(\by^{(0)}) \left( \dfracd{\phi^{(0)}}{t_{1}} \right)^{2},
\end{equation}
where we denoted $\ave{E^{(2)}}$ by $E^{(2)}$ for simplicity.
Finally, we eliminate the unknown amplitudes
from the averaged term $\mathcal{A}$ represented in Eq.~\eqref{eq:averageA}
by substituting Eqs.~\eqref{eq:hypothesis} and \eqref{eq:w}:
\begin{equation}
  \label{eq:A-in-phi0}
  \begin{split}
    & \mathcal{A}(\phi^{(0)})
    = \left[
      \dfrac{E^{(2)}-\Vbend^{(2)}(\phi^{(0)})}{2}
      - \dfrac{1}{4} C^{33}(\by^{(0)}) \left( \dfracd{\phi^{(0)}}{t_{1}} \right)^{2}
    \right] T_{\nu}, \\
  \end{split}
\end{equation}
where
\begin{equation}
  \label{eq:Tnu}
  \begin{split}
    T_{\nu}
    & = - \left[
      \dfrac{M_{1}\sin\phi^{(0)}}{M_{2}-M_{1}\cos\phi^{(0)}} \nu_{1}
      - \dfrac{M_{1}\sin\phi^{(0)}}{M_{2}+M_{1}\cos\phi^{(0)}} \nu_{2}
  \right].
  \end{split}
\end{equation}
We underline that the averaged term $\mathcal{A}$ depends on
the constants $\nu_{1}, \nu_{2}$ and $E^{(2)}$.

\subsection{Final result}

Substituting Eq.~\eqref{eq:A-in-phi0} into Eq.~\eqref{eq:Oepsilon2-averaged},
we have
\begin{equation}
  \label{eq:Oepsilon2-closed}
  \begin{split}
    & \dfracd{{}^{2}\phi^{(0)}}{t_{1}^{2}}
    + F_{\nu}(\phi^{(0)}) \left( \dfracd{\phi^{(0)}}{t_{1}} \right)^{2}
    + G_{\nu}(\phi^{(0)}) = 0 \\
  \end{split}
\end{equation}
where
\begin{equation}
  F_{\nu}(\phi^{(0)})
  = \dfrac{1}{2C^{33}(\by^{(0)})} \dfracp{C^{33}}{\phi}(\by^{(0)})
  + \dfrac{1}{4} T_{\nu},
\end{equation}
\begin{equation}
  \label{eq:Gnu}
  G_{\nu}(\phi^{(0)})
  = \dfrac{1}{C^{33}(\by^{(0)})} \left[
    \dfracd{\Vbend^{(2)}}{\phi}(\phi^{(0)})
    - \dfrac{E^{(2)}-\Vbend^{(2)}(\phi^{(0)})}{2} T_{\nu}
  \right],
\end{equation}
and
\begin{equation}
  C^{33}(\phi^{(0)}) = \dfrac{l_{\ast}^{2}}{2} ( M_{2} - M_{1}\cos\phi^{(0)}).
\end{equation}
Equation \eqref{eq:Oepsilon2-closed} is the closed equation
for the slow bending motion.
It is reproduced as the Euler-Lagrange equation
of the effective Lagrangian
\begin{equation}
  \Leff \left( \phi^{(0)}, \dfracd{\phi^{(0)}}{t_{1}} \right)
  = \dfrac{1}{2} \Meff(\phi^{(0)}) \left( \dfracd{\phi^{(0)}}{t_{1}} \right)^{2}
  - \Veff(\phi^{(0)}).
\end{equation}
Here, the effective (dimensionless) mass $\Meff(\phi^{(0)})$ is
\begin{equation}
  \label{eq:Meff}
  \begin{split}
    & \Meff(\phi^{(0)})
    = \exp \left[ 2 \int_{0}^{\phi^{(0)}} F_{\nu}(z) dz \right] \\
    & = \left( \dfrac{M_{2}-M_{1}\cos\phi^{(0)}}{M_{2}-M_{1}} \right)^{1-\nu_{1}/2}
    \left( \dfrac{M_{2}+M_{1}}{M_{2}+M_{1}\cos\phi^{(0)}} \right)^{\nu_{2}/2},
  \end{split}
\end{equation}
and the effective potential $\Veff(\phi^{(0)})$ is
\begin{equation}
  \label{eq:Veff}
  \Veff(\phi^{(0)}) = \int_{0}^{\phi^{(0)}} \Meff(z) G_{\nu}(z) dz.
\end{equation}
Note that the physical dimension of $\Veff$ differs from $\Vbend^{(2)}$
due to the factor $1/C^{33}$.

The effective potential $\Veff$ of Eq.~\eqref{eq:Veff} is the main product of the theory.
A remarkable observation is that $\Veff$ depends on energy $E^{(2)}$
and the normal mode energy distribution $(\nu_{1},\nu_{2})$
through $\Meff$ [Eq.~\eqref{eq:Meff}] and $G_{\nu}$ [Eq.~\eqref{eq:Gnu}].
Examples of the effective potential are exhibited in Sec.~\ref{sec:effective-potentials}.

\section{Effective potential}
\label{sec:effective-potentials}

We exhibit examples of the effective potential
with varying the parameters $E^{(2)}$ and $\nu_{1}$
(remember $\nu_{2}=1-\nu_{1}$).
The equal mass condition $m_{2}=m$ is assumed unless there is a notice.
Note that the in-phase (anti-phase) mode is
the mode-$1$ (mode-$2$) as defined in Sec.~\ref{sec:explicit-form-A}.

\subsection{Examples without bending potential}

\begin{figure}
  \centering
  \includegraphics[width=8.0cm]{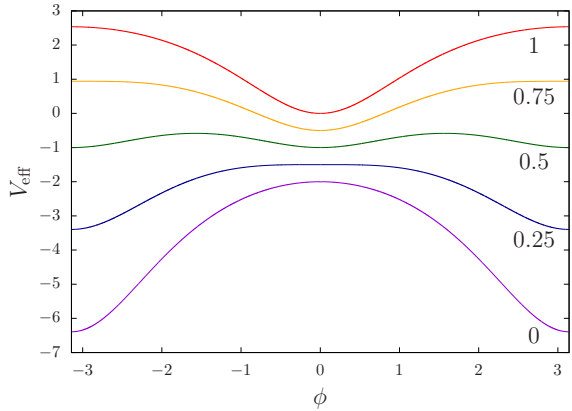}
  \caption{The effective potential $\Veff(\phi)$. $V_{\rm bent}\equiv 0$.
    The numbers in the panels represent the values of $\nu_{1}$,
    while $\nu_{2}=1-\nu_{1}$.
    The total energy $E^{(2)}$ is an overall factor of $\Veff$
    and is set as $E^{(2)}=1$.
    Graphs are shifted in the vertical direction by a graphical reason.
    }
  \label{fig:Veff_noVbend}
\end{figure}

First of all, we observe the effective potential $\Veff$
without bending potential, $\Vbend\equiv 0$,
to observe the simplest case.
We exhibit effective potentials for some values of $\nu_{1}$ ($\nu_{2}=1-\nu_{1}$)
in Fig.~\ref{fig:Veff_noVbend}.
The dynamical contribution is completely opposite
between the in-phase mode and the anti-phase mode.
The in-phase mode makes a valley at $\phi=0$,
while the anti-phase mode makes a valley at $\phi=\pi$.
A precise analysis reveals that there are the two local minima at $\phi=0$ and $\pi$
in the interval of $\nu_{1}\in (1/4, 3/4)$.
The coexistence interval is generalized to
\begin{equation}
  \nu_{1} \in \left( \dfrac{M_{2}-M_{1}}{2M_{2}}, \dfrac{M_{2}+M_{1}}{2M_{2}} \right)
\end{equation}
for any value of $m_{2}$. See Appendix \ref{sec:Zmu-analysis} for details.

\subsection{Examples with a bending potential}

\begin{figure}
  \centering
  \includegraphics[width=8.0cm]{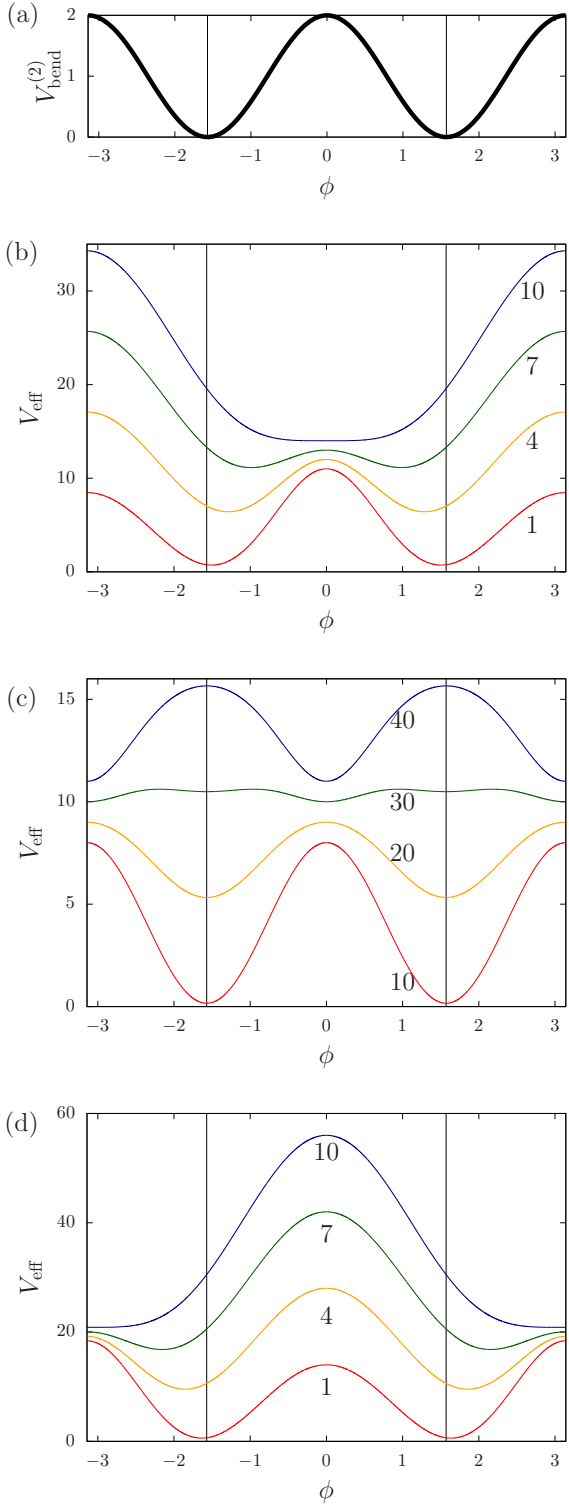}
  \caption{The bending potential $\Vbend^{(2)}(\phi)=2\cos\phi+1$ (a)
    and the effective potential $\Veff(\phi)$ [(b)-(d)].
    (b) $(\nu_{1},\nu_{2})=(1,0)$ (the in-phase mode).
    (c) $(\nu_{1},\nu_{2})=(0.5,0.5)$ (a mixed mode).
    (d) $(\nu_{1},\nu_{2})=(0,1)$ (the anti-phase mode).
    The numbers in the panels (b)-(d) represent values of $E^{(2)}$.
    Graphs are shifted in the vertical direction by a graphical reason.
    The black vertical straight lines mark the minimum points of $\Vbend^{(2)}(\phi)$.
    }
  \label{fig:Veff}
\end{figure}

Next, we introduce an example of the bending potential as
\begin{equation}
  \label{eq:Vbend-example}
  \Vbend^{(2)}(\phi) = \cos 2\phi + 1.
\end{equation}
This potential has the two minima at $\phi=\pm\pi/2$.
We set the equal mass condition, $m_{2}=m$.
Since $\Veff$ depends on the normal mode energy distribution $(\nu_{1},\nu_{2})$
and the total energy $E^{(2)}$,
we show graphs of the effective potential for $(\nu_{1},\nu_{2})=(1,0)$ (in-phase),
$(1/2,1/2)$ (mixed), and $(0,1)$ (anti-phase)
with varying the value of $E^{(2)}$ in Fig.~\ref{fig:Veff}.
The effective potential $\Veff$ is similar to the bending potential $\Vbend^{(2)}$
when the total energy $E^{(2)}$ is small.
As the total energy increases, the local minimum points move from $\phi=\pm\pi/2$
towards $\phi=0$ and/or $\phi=\pi$.
The local minimum points of $\Veff$ are the dynamically induced conformations
(DICs).

\begin{figure}
  \centering
  \includegraphics[width=8.0cm]{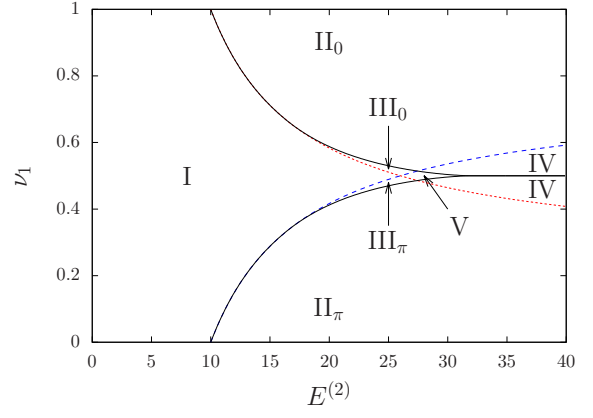}
  \caption{The phase diagram on the $(E^{(2)},\nu_{1})$ plane
    with the equal mass condition $m_{2}=m$.
    $m=1$. $\nu_{2}=1-\nu_{1}$.
    The effective potential takes a local minimum
    at $\phi=0$ over the red dotted line,
    and at $\phi=\pi$ under the blue dashed line,
    where the two lines are obtained
    from Eqs.~\eqref{eq:0-stable} and \eqref{eq:pi-stable} respectively.
    $\phi=\phi_{\sharp}~(\neq 0,\pi)$  is a local minimum point
    between the two solid black lines.
    See Table \ref{tab:combination-localminima} for the types from I to V.
  }
  \label{fig:PhaseDiagramEqualMass}
\end{figure}

The effective potential is determined at each point on the $(E^{(2)},\nu_{1})$ plane,
and yields the set of the local minimum points.
We categorize the local minimum points into the three classes:
$\phi=0, \pi$, and $\phi_{\sharp}(\neq 0,\pi)$.
The three classes induce the seven types of sets
as arranged in Table \ref{tab:combination-localminima}.
By using the seven types, the $(E^{(2)},\nu_{1})$ plane is divided into regions
each of which is assigned by a type of the set
as reported in Fig.~\ref{fig:PhaseDiagramEqualMass}.
We stress that the seven types are realized by changing the total energy $E^{(2)}$
and the mode energy distribution $\nu_{1}$.

\begin{table}[htb]
  \centering
  \caption{The seven types of local minimum point sets of the effective potential $\Veff$.
    The symbol $\phi_{\sharp}$ represents a conformation
    which is neither $\phi=0$ nor $\phi=\pi$.
    For each type the conformation $\phi$ with the symbol M (O)
    is a local minimum point (not a local minimum point).}
  \begin{tabular}{c|ccccccc}
    \hline
    Conformation $\phi$ & I & II${}_{0}$ & II${}_{\pi}$ & III${}_{0}$ & III${}_{\pi}$ & IV & V \\
    \hline
    $\phi_{\sharp}(\neq 0,\pi)$ & M & O & O & M & M & O & M \\
    $0$ & O & M & O & M & O & M & M \\
    $\pi$ & O & O & M & O & M & M & M \\
    \hline
  \end{tabular}
  \label{tab:combination-localminima}
\end{table}

\begin{figure}
  \centering
  \includegraphics[width=8cm]{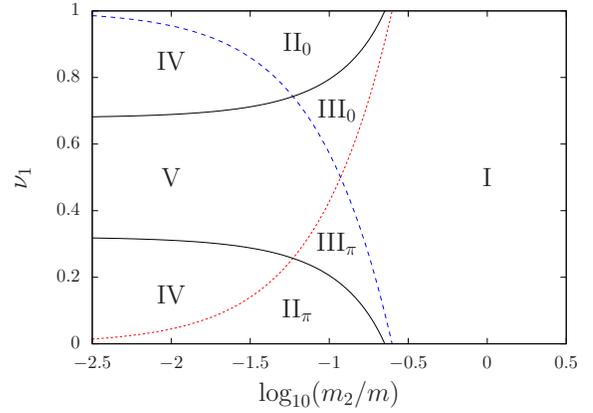}
  \caption{
    The phase diagram on the plane $(m_{2}/m,\nu_{1})$ with $E^{(2)}=4$.
    The meanings of the lines are the same as Fig.~\ref{fig:PhaseDiagramEqualMass}.
    See Table \ref{tab:combination-localminima} for the types from I to V.
  }
  \label{fig:PhaseDiagram}
\end{figure}

Finally, we present a phase diagram by varying the center mass $m_{2}$
with fixing $E^{(2)}=4$ in Fig.~\ref{fig:PhaseDiagram}.
We give two remarks.
First, the seven types are also realized by changing the center mass $m_{2}$.
The regions assigned by the types III${}_{0}$, III${}_{\pi}$, IV, and V
are enhanced comparing with Fig.~\ref{fig:PhaseDiagramEqualMass}.
This fact suggests that the mass distribution is useful to control the conformation. 
Second, the dynamical contribution dominates the effective potential
when $m_{2}$ is small.
This domination can be explained from Eq.~\eqref{eq:Tnu},
which is a part of the averaged term $\mathcal{A}$,
and Eq.~\eqref{eq:M2M1}, which is the definitions of $M_{2}$ and $M_{1}$.
We have $M_{2}\to M_{1}$ as $m_{2}\to 0$ from Eq.~\eqref{eq:M2M1}.
Thus, the denominators of the function $T_{\nu}$ can be close to $0$
near $\phi^{(0)}=0,\pi$,
and hence contribution from the averaged term becomes large.

\section{Numerical tests}
\label{sec:numerics}

We verify efficiency of the effective potential through numerical
simulations of the model.

\subsection{Setting}
\label{sec:setting}
Numerical simulations are performed by using the fourth order
symplectic integrator \cite{yoshida-90}
for the Hamiltonian
\begin{equation}
  \label{eq:model-Ham}
  H = \dfrac{1}{2} \sum_{j=1}^{3} \dfrac{\norm{\bp_{j}}^{2}}{m_{j}}
  + V(\br_{1},\br_{2},\br_{3}),
  \quad
  \bp_{j} = m_{j} \dot{\br}_{j},
\end{equation}
which is the Legendre transform of Eq.~\eqref{eq:model}.
The time step is set as $\Delta t=10^{-3}$.
The relative energy error is suppressed in the reported simulations
as $|(E_{\rm num}-E_{0})/E_{0}|<10^{-10}$,
where $E_{0}$ and $E_{\rm num}$ are respectively
the initial energy and the numerically obtained energy.

The two springs are assumed to be linear and $\Vspring$ is
\begin{equation}
  \Vspring(l_{1},l_{2}) = \dfrac{k}{2} \left[
    (l_{1}-l_{\ast})^{2} + (l_{2}-l_{\ast})^{2} \right],
\end{equation}
because the theory includes only the linear part of the springs.
The bending potential is $\Vbend=\epsilon^{2}\Vbend^{(2)}$,
and we use Eq.~\eqref{eq:Vbend-example} as $\Vbend^{(2)}$.
The small parameter $\epsilon$ is fixed as $\epsilon=0.1$.
The masses are equal and $m_{1}=m_{2}=m_{3}=m=1$.
The spring constant is $k=10$,
and the natural length is $l_{\ast}=1$.

\subsection{Initial condition}
\label{sec:initial-condition}
We set the initial condition as follows.
All the beads have zero initial velocities in the $x$- and $y$-directions.
The beads are once placed at the natural lengths of the springs
with the bending angle $\phi_{\ast}$.
Then, under the hypothesis ({\it H}),
we give small displacements of $l_{1},l_{2}$, and $\phi$
so as to excite the normal modes of the springs for a given pair of $(\nu_{1},\nu_{2})$.
The initial condition, denoted by the subscript $0$, is summarized as
\begin{equation}
  \label{eq:initial-condition}
  \left\{
    \begin{split}
      & 
      \begin{pmatrix}
        l_{1,0} \\ l_{2,0} \\ \phi_{0}
      \end{pmatrix}
      =
      \begin{pmatrix}
        l_{\ast} \\ l_{\ast} \\ \phi_{\ast}
      \end{pmatrix}
      + \epsilon w \left[
        \sqrt{\nu_{1}}
        \bp_{\rm in}(\by^{(0)})
        + \sqrt{\nu_{2}}
        \bp_{\rm anti}
      \right], 
      \\
      & 
      \begin{pmatrix}
        \dot{l}_{1,0} \\ \dot{l}_{2,0} \\ \dot{\phi}_{0} \\
      \end{pmatrix}
      =
      \begin{pmatrix}
        0 \\ 0 \\ 0 \\
      \end{pmatrix}.
    \end{split}
  \right.
\end{equation}
See Eq.~\eqref{eq:P} for the definitions of $\bp_{\rm in}$ and $\bp_{\rm anti}$.
We note that the amplitude $w$ is of $O(\epsilon^{0})$.
In the Hamiltonian system of Eq.~\eqref{eq:model-Ham},
a corresponding initial condition is
\begin{equation}
  \left\{
    \begin{split}
      & \br_{1,0} =
      \begin{pmatrix}
        - l_{1,0}\cos(\phi_{0}/2) \\ l_{1,0} \sin(\phi_{0}/2)
      \end{pmatrix},
      ~
      \br_{3,0} =
      \begin{pmatrix}
        l_{2,0}\cos(\phi_{0}/2) \\ l_{2,0} \sin(\phi_{0}/2)
      \end{pmatrix}
      \\
      & \br_{2,0} =\bp_{1,0} = \bp_{2,0} = \bp_{3,0} = \bzero,
    \end{split}
  \right.
\end{equation}
where $\bzero$ is the two-dimensional zero vector.
The above initial condition gives the second order total energy as
\begin{equation}
  \label{eq:E2-example}
  E^{(2)} = \dfrac{k}{2} w^{2} + \Vbend^{(2)}(\phi_{0}).
\end{equation}
In the next section we set $\phi_{\ast}=\pi/2$,
which is a local minimum point of the bending potential $\Vbend$.
The initial condition Eq.~\eqref{eq:initial-condition}
hence has two free parameters of the amplitude $w$,
which is equivalent with $E^{(2)}$ through Eq.~\eqref{eq:E2-example},
and the normal mode energy ratio $\nu_{1}$ (remember $\nu_{2}=1-\nu_{1}$).

\subsection{Efficiency of the effective potential}

\begin{figure}[t]
  \centering
  \includegraphics[width=8.5cm]{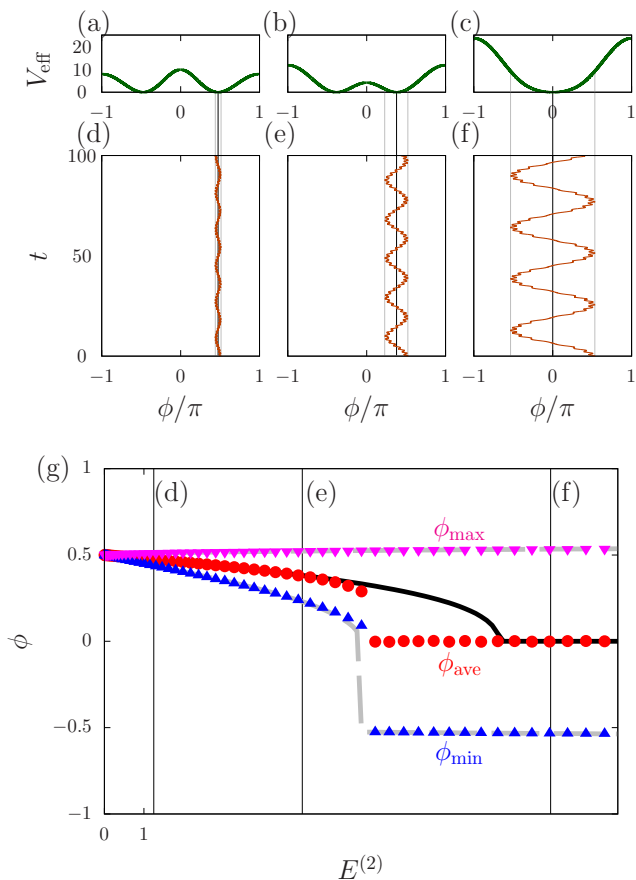}
  \caption{
    Graphs of the effective potential [(a)-(c)] and temporal evolution of $\phi(t)$
    with the reference conformation $\phi_{\ast}=\pi/2$ [(d)-(f)].
    $(\nu,_{1},\nu_{2})=(1,0)$.
    The amplitude $w$ is (a,d) $w=0.5$, (b,e) $w=1$, and (c,f) $w=1.5$.
    The vertical black straight solid line represents the minimum point of $\Veff$.
    The two vertical gray straight dashed lines represent
    the two values of $\phi$ satisfying $\Veff(\phi)=\Veff(\phi_{0})$
    and belonging to the same valley.
    (g) Energy dependence of the time average $\phi_{\rm ave}$ (red circles)
    with the minimum $\phi_{\rm min}$ (blue triangles)
    and the maximum $\phi_{\rm max}$ (purple inverse triangles)
    of $\phi$ in the time interval $t\in [0,1000]$.
    The black solid and the gray dashed thick lines corresponds respectively
    to the vertical black and gray straight lines presented in the panels (d), (e), and (f).
    The three black vertical solid lines indicate the values of energy
    which correspond to the panels (d), (e), and (f) from left to right.
  }
  \label{fig:phi_phiinipiover2}
\end{figure}

We concentrate on the in-phase mode, $(\nu_{1},\nu_{2})=(1,0)$.
Temporal evolution of $\phi(t)$ is exhibited
in Figs.~\ref{fig:phi_phiinipiover2}(d), (e), and (f)
for three values of $w$, corresponding to three values of $E^{(2)}$
[see Eq.~\eqref{eq:E2-example}].
Small and fast oscillation of $\phi(t)$ comes from $\phi^{(1)}$
and is induced from $l_{j}^{(1)}$,
which are governed by Eq.~\eqref{eq:Oepsilon1}.
When the amplitude $w$ is small,
the bending angle $\phi$ almost stays around the initial value $\phi_{0}$
[Fig.~\ref{fig:phi_phiinipiover2}(d)],
as it is predicted from the bending potential $\Vbend$.
However, the amplitude of oscillation of $\phi$ becomes large as $w$ gets large,
and the center of oscillation approaches to the zero
[Figs.~\ref{fig:phi_phiinipiover2}(e) and (f)].
We estimate the center of oscillation by the time average
\begin{equation}
  \label{eq:phi-center}
  \phi_{\rm ave} = \dfrac{1}{T} \int_{0}^{T} \phi(t) dt,
  \quad
  T=1000.
\end{equation}
The estimated center is plotted as a function of $E^{(2)}$
in Fig.~\ref{fig:phi_phiinipiover2}(g)
with the minimum $\phi_{\rm min}$ and the maximum $\phi_{\rm max}$
of $\phi(t)$ in $t\in[0,1000]$.

Two remarks are in order.
First, the minimum and the maximum of $\phi$
are well predicted by $\Veff$.
There is a gap between $\phi_{\rm ave}$ and the bottom of $\Veff$
in the energy interval approximated by $E^{(2)}\in [6.8,10]$.
The gap is not a counterevidence but a supporting evidence of the theory.
This gap comes from the inequality $\Veff(0)<\Veff(\phi_{0})$,
which implies that the bending angle climbs over the saddle point at $\phi=0$.
This passing is confirmed by the jump of the theoretical minimum value of $\phi$.
Second, the center of oscillation is continuously modified as the spring energy increases.
The conformation is determined
not by a local minimum of the bending potential $\Vbend$
but by a local minimum of the effective potential $\Veff$ derived from dynamics.
We conclude that the effective potential successfully predicts
the slow bending motion.

We provide movies in Supplemental Material \cite{SM}
to show dynamics of the system.
See Appendix \ref{sec:movie} for explanation on the movies.

\section{Summary and discussions}
\label{sec:summary}

We have investigated the three-body bead-spring model
and demonstrated the dynamically induced conformation (DIC).
The fast motion of the springs induces the effective potential
for the slow bending motion,
and the conformation is governed by the bottoms
of the effective potential instead of the bending potential.
One crucial remark is that the effective potential
depends on the excited normal modes and its energy:
The effective potential tends to have the minimum at $\phi=0$ ($\phi=\pi$),
by exciting the in-phase (anti-phase) mode.
Moreover, a mixed mode makes the two local minima
at $\phi=0$ and $\pi$.

We have developed a theory to derive the effective potential
based on the multiple-scale analysis and the averaging method.
The main idea of the theory is to introduce a hypothesis
inspired from the adiabatic invariant.
The hypothesis with the energy conservation law
eliminate the unknown variables being unavoidable in autonomous systems,
and the elimination introduces the mode dependence
and the total energy dependence into the effective potential.
A theory for a generic system can be found in
Refs.~\cite{yamaguchi-etal-21,yamaguchi-22}.

Efficiency of the theoretically obtained effective potential is successfully examined
through numerical simulations.
The bending angle oscillates in general, and the center of oscillation
is not a bottom of the bending potential,
but a bottom of the effective potential.
An extreme example is that a local maximum point of the bending potential
becomes a local minimum point of the effective potential.
The amplitude of the bending angle oscillation is also predicted
by the effective potential.

The studied model is quite simple as we neglected
the excluded volume effect, for instance.
The potential of the excluded volume effect,
and any other potentials between the two beads of the ends,
can be treated in the same way as the bending potential discussed in this article
(see Appendix \ref{sec:GeneralPotential}).
Therefore, the phenomenon of DIC is universal as long as
the two assumptions {\it (A1)} and {\it (A2)} hold.

We give three discussions on {\word} revealed in this article:
universality, application to control, and the beat effect.
First, {\word} must be universal
since the essential mechanism to have {\word} is existence of multiple timescales.
Indeed, numerical simulations show that $N$-body bead-spring systems exhibit {\word}
\cite{yanagita-konishi-jp}.
Details will be reported elsewhere.
Second, it is interesting to use {\word} 
for controlling the conformation of proteins by changing energy.
Control of a robot is also an interesting subject
by changing the center mass $m_{2}$ as shown in Fig.~\ref{fig:PhaseDiagram}.
Finally, we have neglected the beat effect between the eigenfrequencies
of the fast springs, namely $\lambda_{1}=\lambda_{2}$ at $\phi=\pm\pi/2$.
The beat effect may trigger the Arnold diffusion \cite{arnold-64}
since the full dynamics has the three degrees of freedom, $(l_{1},l_{2},\phi)$
(see, for example, Refs.~\cite{manikandan-keshavamurthy-14,firmbach-etal-18}
for the recent progress on systems of three degrees of freedom).
It will be interesting to observe evolution of the system
in a very long time beyond the slow timescale $t_{1}=\epsilon t$.

\acknowledgements
Y.Y.Y. acknowledges the support of JSPS KAKENHI Grant Numbers 16K05472 and 21K03402.
T.Y. acknowledges the support of JSPS KAKENHI Grant Numbers 18K03471 and 21K03411.
T.K. is supported by Chubu University Grant (A).
M.T. is supported by the Research Program of "Dynamic Alliance for Open Innovation Bridging Human, Environment and Materials" in "Network Joint Research Center for Materials and Devices",
and a Grant-in-Aid for Scientific Research (C) ( No. 22654047, 25610105, and 19K03653 ) from JSPS.
The authors express their thanks to the anonymous referees
who suggested to input the bending potential.

\appendix

\section{Lagrangians of the three-body bead-spring model }
\label{sec:three-body-Rouse-model}

The system of Eq.~\eqref{eq:model} has the two-dimensional translational symmetry
and the rotational symmetry.
We reduce Eq.~\eqref{eq:model} and derive Eq.~\eqref{eq:Lagrangian-y}
by introducing the internal coordinates.
For the reduction we perform three changes of variables.

The first change of variables introduces the vectors along the springs,
denoted by $\bq_{1}$ and $\bq_{2}$, with the center-of-mass $\bq_{\rm G}$.
This change of variables is expressed as
\begin{equation}
  \begin{pmatrix}
    \bq_{1} \\ \bq_{2} \\ \bq_{\rm G}
  \end{pmatrix}
  =
  \begin{pmatrix}
    -1 & 1 & 0 \\
    0 & -1 & 1 \\
    m/M & m_{2}/M & m/M \\
  \end{pmatrix}
  \begin{pmatrix}
    \br_{1} \\ \br_{2} \\ \br_{3}
  \end{pmatrix}
\end{equation}
with the total mass
\begin{equation}
  M=2m+m_{2}.
\end{equation}
Since each element of $\bq_{\rm G}$ is a cyclic coordinate by an assumption
and $\dot{\bq}_{\rm G}$ is conserved,
we set $\dot{\bq}_{\rm G}\equiv\boldsymbol{0}$ without loss of generality.
This setting reduces $\bq_{\rm G}$ and $\dot{\bq}_{\rm G}$
from the Lagrangian, which is written as
\begin{equation}
  \label{eq:Lagrangian-q}
  L = \dfrac{1}{2} \sum_{i,j=1}^{2} A^{ij} \dot{\bq}_{i} \cdot \dot{\bq}_{j}
  - V(\bq_{1},\bq_{2}),
\end{equation}
where we assumed that the potential energy function $V$ depends on
only $\bq_{1}$ and $\bq_{2}$.
$A^{ij}$ is the $(i,j)$ element of the size-$2$ matrix $\mA$, which is defined by
\begin{equation}
  \label{eq:matrix-A}
  \mA =
  \begin{pmatrix}
    M_{2} & M_{1} \\
    M_{1} & M_{2} \\
  \end{pmatrix},
  \quad
  M_{2} = \dfrac{m(m+m_{2})}{M},
  \quad
  M_{1} = \dfrac{m^{2}}{M}.
\end{equation}

The second change of variables introduces the polar coordinates
$(l_{j},\theta_{j})$, where $l_{j}$ is the length of $\bq_{j}$
and $\theta_{j}$ is the angle of $\bq_{j}$ measured from
a fixed direction on $\mathbb{R}^{2}$.
The vectors $\bq_{j}$ and $\dot{\bq}_{j}$ are then written as
\begin{equation}
  \bq_{j} = l_{j} \be_{rj},
  \quad
  \dot{\bq}_{j} = \dot{l}_{j} \be_{rj} + l_{j}\dot{\theta}_{j} \be_{\theta j}
\end{equation}
where $\be_{rj}$ is the unit vector to the radial direction of $\bq_{j}$,
and $\be_{\theta j}$ is the unit vector to the angle direction.

As the third change of variables, we define
\begin{equation}
  \phi = \theta_{2} - \theta_{1},
  \quad
  \psi = \theta_{2} + \theta_{1},
\end{equation}
where $\phi$ represents the bending angle (see Fig.~\ref{fig:Rouse3model}).
The variables $l_{1},l_{2},\phi,$ and $\psi$ describes the Lagrangian 
\begin{equation}
  \label{eq:Lagrangian-z}
  L = \dfrac{1}{2} \sum_{\alpha,\beta=1}^{4} B^{\alpha\beta}(\bz)
  \dot{z}_{\alpha} \dot{z}_{\beta}
  - V(l_{1},l_{2},\phi)
\end{equation}
where $\bz=(z_{1},z_{2},z_{3},z_{4})=(l_{1},l_{2},\phi,\psi)$
and $V$ does not depend on $\psi$ by the assumption of rotational symmetry.
The four-dimensional vector is represented by $\bz\in\mathbb{R}^{4}$
to distinguish from the three-dimensional vector $\by\in\mathbb{R}^{3}$
used in Eq.~\eqref{eq:Lagrangian-y}.
$B^{\alpha\beta}$ is the $(\alpha,\beta)$ element of the size-$4$ matrix $\mB$.
The matrix $\mB$ is symmetric,
and we show only the upper triangle elements.
The diagonal elements are
\begin{equation}
  \label{eq:B-diagonal}
  \left\{
    \begin{split}
      & B^{11} = B^{22} = M_{2}, \\
      & B^{33} = \dfrac{1}{4}M_{2}(l_{1}^{2}+l_{2}^{2}) - \dfrac{1}{2}M_{1}l_{1}l_{2}\cos\phi, \\
      & B^{44} = \dfrac{1}{4}M_{2}(l_{1}^{2}+l_{2}^{2}) + \dfrac{1}{2}M_{1}l_{1}l_{2}\cos\phi, \\
    \end{split}
  \right.
\end{equation}
and the off-diagonal elements are
\begin{equation}
  \label{eq:B-offdiagonal}
  \left\{
    \begin{split}
      & B^{12} = M_{1}\cos\phi, \\
      & B^{13} = B^{14} = -\dfrac{1}{2}M_{1}l_{2}\sin\phi, \\
      & B^{23} = - B^{24} = -\dfrac{1}{2}M_{1}l_{1}\sin\phi, \\
      & B^{34} = - \dfrac{1}{4}M_{2}(l_{1}^{2}-l_{2}^{2}). \\
    \end{split}
  \right.
\end{equation}

The Lagrangian of Eq.~\eqref{eq:Lagrangian-z} does not depend on $\psi$
and $\psi$ is a cyclic coordinate.
The conjugate momentum $p_{\psi}$,
which corresponds to the total angular momentum, is defined by
\begin{equation}
  \label{eq:ppsi-definition}
  p_{\psi} = \dfracp{L}{\dot{z}_{4}} = \sum_{\alpha=1}^{4} B^{4\alpha} \dot{z}_{\alpha}
\end{equation}
and is conserved.
Eliminating $\dot{z}_{4}(=\dot{\psi})$ from the kinetic energy,
we obtain the Lagrangian
\begin{equation}
    L
    = \dfrac{1}{2} \sum_{\alpha,\beta=1}^{3} C^{\alpha\beta}(\by) \dot{y}_{\alpha} \dot{y}_{\beta}
  + \dfrac{[p_{\psi}(\by,\dot{\by})]^{2}}{2B^{44}(\by)}
  - V(\by),
\end{equation}
where we identified $\mB(\bz)$ and $\mB(\by)$
since $\mB$ does not depend on $\psi$.
Assuming $p_{\psi}=0$, we obtain Eq.~\eqref{eq:Lagrangian-y}
because contribution from $p_{\psi}$ to the Euler-Lagrange equation vanishes.
The $(\alpha,\beta)$ element of the size-$3$ symmetric matrix $\mC$ is defined by
\begin{equation}
  C^{\alpha\beta}(\by)
  = B^{\alpha\beta}(\by) - \dfrac{1}{B^{44}(\by)} B^{4\alpha}(\by) B^{4\beta}(\by).
\end{equation}
The diagonal elements are
\begin{equation}
  \label{eq:C-diagonal}
  \left\{
    \begin{split}
      C^{11}(\bx)
      & = M_{2}
      - \dfrac{M_{1}^{2}l_{2}^{2}\sin^{2}\phi}
      {M_{2}(l_{1}^{2}+l_{2}^{2})+2M_{1}l_{1}l_{2}\cos\phi},
      \\
      C^{22}(\bx)
      & = M_{2}
      - \dfrac{M_{1}^{2}l_{1}^{2}\sin^{2}\phi}
      {M_{2}(l_{1}^{2}+l_{2}^{2})+2M_{1}l_{1}l_{2}\cos\phi},
      \\
      C^{33}(\bx)
      & = \dfrac{1}{4}M_{2}(l_{1}^{2}+l_{2}^{2})
      -\dfrac{1}{2}M_{1}l_{1}l_{2}\cos\phi \\
      & - \dfrac{M_{2}^{2}(l_{1}^{2}-l_{2}^{2})^{2}}{4M_{2}(l_{1}^{2}+l_{2}^{2})+8M_{1}l_{1}l_{2}\cos\phi},
    \end{split}
  \right.
\end{equation}
and the off-diagonal elements are
\begin{equation}
  \label{eq:C-offdiagonal}
  \left\{
    \begin{split}
      & C^{12}(\bx) 
      = M_{1}\cos\phi
      + \dfrac{M_{1}^{2}l_{1}l_{2}\sin^{2}\phi} 
      {M_{2}(l_{1}^{2}+l_{2}^{2})+2M_{1}l_{1}l_{2}\cos\phi},
      \\
      & C^{13}(\bx) 
      = -\dfrac{1}{2}M_{1}l_{2}\sin\phi
      - \dfrac{\frac{1}{2}M_{1}M_{2}(l_{1}^{2}-l_{2}^{2})l_{2}\sin\phi}
      {M_{2}(l_{1}^{2}+l_{2}^{2})+2M_{1}l_{1}l_{2}\cos\phi},
      \\
      & C^{23}(\bx) 
      = - \dfrac{1}{2}M_{1}l_{1}\sin\phi
      + \dfrac{\frac{1}{2}M_{1}M_{2}(l_{1}^{2}-l_{2}^{2})l_{1}\sin\phi}
      {M_{2}(l_{1}^{2}+l_{2}^{2})+2M_{1}l_{1}l_{2}\cos\phi}.
      \\
    \end{split}
  \right.
\end{equation}

\section{Kapitza pendulum}
\label{eq:Kapitza-pendulum}
We review an analysis of the Kapitza pendulum.
This review is adjusted to our theory
for easily capturing a road map of long computations.

We consider a pendulum on the $xy$ plane
where the $y$ axis points to the upward direction of the gravity $g$.
The pendulum has the length $l$ and a point mass $m$ at the tip.
The angle $\phi$ is taken from the downward direction of the $y$ axis
to the anti-clockwise direction.
An external force oscillates the pivot of the pendulum
along the $y$ axis with the amplitude $a$ and the frequency $\omega$.
The position $(x,y)$ of the point mass is then written as
\begin{equation}
  x = l \sin\phi, \quad y =  -l\cos\phi - a \cos(\omega t+\delta),
\end{equation}
where $\delta$ is the initial phase of the pivot.
Constructing the Lagrangian, we have the Euler-Lagrange equation
for $\phi$ as
\begin{equation}
  \label{eq:Kapitza-EL}
  \dfracd{{}^{2}\phi}{\bar{t}^{2}}
  = - \left[
    \left( \dfrac{\omega_{0}}{\omega} \right)^{2}
    + \dfrac{a}{l} \cos (\bar{t}+\delta)
  \right]
  \sin\phi,
\end{equation}
where $\omega_{0}=\sqrt{g/l}$ and $\bar{t}=\omega t$.
If no external oscillation is applied to the pivot, namely $a=0$,
the unique stable stationary point is clearly $\phi=0$.

We assume that (i) the amplitude $a$ of the oscillating pivot
is much smaller than the pendulum length $l$ and is of $O(\epsilon)$,
(ii) the frequency $\omega_{0}$ is much smaller than $\omega$
and is of $O(\epsilon)$,
where $\epsilon$ is a dimensionless small parameter.
These assumptions imply 
\begin{equation}
  \label{eq:Kapitza-scaling}
  \dfrac{a}{l} = \epsilon \alpha,
  \quad
  \dfrac{\omega_{0}}{\omega} = \epsilon \beta,
  \quad
  |\epsilon|\ll 1,
\end{equation}
where $\alpha$ and $\beta$ are of $O(\epsilon^{0})$.

We introduce two timescales of
\begin{equation}
  t_{0} = \bar{t}, \quad t_{1} = \epsilon \bar{t},
\end{equation}
which induce
\begin{equation}
  \dfracd{}{\bar{t}} = \dfracp{}{t_{0}} + \epsilon \dfracp{}{t_{1}}.
\end{equation}
The angle $\phi$ is also expanded as
\begin{equation}
  \phi(t) = \phi^{(0)}(t_{1}) + \epsilon \phi^{(1)}(t_{0},t_{1}).
\end{equation}
Substituting the above expansions into the Euler-Lagrange equation,
Eq.~\eqref{eq:Kapitza-EL}, we have the expanded equation
\begin{equation}
  \label{eq:Kapitza-EL-expanded}
  \begin{split}
    & \epsilon^{2} \dfracp{{}^{2}\phi^{(0)}}{t_{1}^{2}}
    + \epsilon \dfracp{{}^{2}\phi^{(1)}}{t_{0}^{2}}
    + 2\epsilon^{2} \dfrac{\partial^{2}\phi^{(1)}}{\partial t_{0}\partial t_{1}}
    + \epsilon^{3} \dfracp{{}^{2}\phi^{(1)}}{t_{1}^{2}}\\
    & = - \left[ \epsilon^{2} \beta^{2} + \epsilon\alpha\cos(t_{0}+\delta) \right]
    \sin(\phi^{(0)}+\epsilon\phi^{(1)}).
  \end{split}
\end{equation}

The equation to $O(\epsilon)$ is
\begin{equation}
  \dfracp{{}^{2}\phi^{(1)}}{t_{0}^{2}} = - \alpha \cos(t_{0}+\delta) \sin\phi^{(0)}(t_{1}).
\end{equation}
Solving the above equation with avoiding secular terms, we have
\begin{equation}
  \label{eq:Kapitza-order1-sol}
  \phi^{(1)}(t_{0},t_{1}) = \alpha \cos(t_{0}+\delta) \sin\phi^{(0)}(t_{1}).
\end{equation}

The equation to $O(\epsilon^{2})$ is
\begin{equation}
  \dfracp{{}^{2}\phi^{(0)}}{t_{1}^{2}}
  + 2 \dfrac{\partial^{2}\phi^{(1)}}{\partial t_{0}\partial t_{1}}
  = - \left[ \beta^{2} \sin\phi^{(0)} + \alpha \phi^{(1)}\cos(t_{0}+\delta) \cos\phi^{(0)} \right].
\end{equation}
Substituting the $O(\epsilon)$ solution Eq.~\eqref{eq:Kapitza-order1-sol}
and averaging over the fast timescale $t_{0}$, we have
\begin{equation}
  \dfracp{{}^{2}\phi^{(0)}}{t_{1}^{2}}
  = - \left( \beta^{2} \sin\phi^{(0)} + \dfrac{\alpha^{2}}{4} \sin 2 \phi^{(0)} \right) .
\end{equation}
The effective potential $V_{\rm eff}(\phi^{(0)})$ satisfying
\begin{equation}
  \dfracp{{}^{2}\phi^{(0)}}{t_{1}^{2}} = - \dfracd{V_{\rm eff}}{\phi^{(0)}}(\phi^{(0)})
\end{equation}
is then obtained as
\begin{equation}
  V_{\rm eff}(\phi^{(0)})
  = - \left( \beta^{2} \cos\phi^{(0)} + \dfrac{\alpha^{2}}{8} \cos 2\phi^{(0)} \right).
\end{equation}
This effective potential has a local minimum at $\phi^{(0)}=\pi$
for $\alpha^{2}>2\beta^{2}$ in addition to $\phi^{(0)}=0$.
The inverted pendulum (i.e. $\phi=\pi$) is therefore stabilized by
sufficiently fast oscillation (i.e. small $\beta$) of the pivot
irrespective of the initial phase $\delta$.

\section{Order of the bending potential}
\label{sec:Vbend-Oepsilon2}

We show $\Vbend^{(0)}(\phi)\equiv 0$ and $\Vbend^{(1)}(\phi)\equiv 0$
under the assumptions ({\it A1}) and ({\it A2}).

\subsection{$O(\epsilon^{0})$}
There is no time derivative term in $O(\epsilon^{0})$,
and we have
\begin{equation}
  \label{eq:AppC-epsilon0}
  \dfracp{\Vbend^{(0)}}{\phi}(\phi^{(0)}) = 0.
\end{equation}
The zeroth order term $\Vbend^{(0)}$ is hence constant,
and we can set $\Vbend^{(0)}(\phi)\equiv 0$ without loss of generality.
We note that the identical zero is induced
because $\phi^{(0)}$ in Eq.~\eqref{eq:AppC-epsilon0} is a variable.
The spring potential $\Vspring$ is not identically zero in general
because it is required to hold in $O(\epsilon^{0})$
\begin{equation}
  \label{eq:AppC-epsilon0-spring}
  \dfracp{\Vspring}{l_{j}}(l_{\ast},l_{\ast}) = 0
  \quad (j=1,2)
\end{equation}
at the point $(l_{1},l_{2})=(l_{\ast},l_{\ast})$.

\subsection{$O(\epsilon)$}
The terms of $O(\epsilon)$ constructs
\begin{equation}
  \mC(\by^{(0)})
  \dfracp{{}^{2}}{t_{0}^{2}}
  \begin{pmatrix}
    l_{1}^{(1)} \\
    l_{2}^{(1)} \\
    \phi^{(1)} \\
  \end{pmatrix}
  =
  \begin{pmatrix}
    -k l_{1}^{(1)} \\
    -k l_{2}^{(1)} \\
    - \left( \partial \Vbend/\partial \phi \right)^{(1)} \\
  \end{pmatrix},
\end{equation}
where
  \begin{equation}
    \label{eq:AppC-epsilon1}
  \left( \dfracp{\Vbend}{\phi} \right)^{(1)}
  = \dfracp{{}^{2}\Vbend^{(0)}}{\phi^{2}}(\phi^{(0)}) \phi^{(1)}
  + \dfracp{\Vbend^{(1)}}{\phi}(\phi^{(0)}).
\end{equation}
The first term of the right-hand side in Eq.~\eqref{eq:AppC-epsilon1}
is zero since $\Vbend^{(0)}\equiv 0$,
and there is no restoring force for the variable $\phi^{(1)}$,
while Eq.~\eqref{eq:AppC-epsilon0-spring} does not imply
the zero restoring force for the springs.
The second term $(\partial \Vbend^{(1)}/\partial\phi)(\phi^{(0)})$
is constant in the timescale $t_{0}$.
If the second term is not zero, $\phi^{(1)}$ has a secular term,
and the secular term breaks the perturbation expansion Eq.~\eqref{eq:expansion-space},
which assumes $|\phi^{(0)}| \gg |\epsilon\phi^{(1)}|$.
Therefore, the second term must be zero and we can set $\Vbend^{(1)}\equiv 0$
without loss of generality.

\section{Matrices in the equations of $O(\epsilon)$}
\label{sec:matrices-Oepsilon1}

We give the explicit forms of the matrix $\mX(\by^{(0)})$
appearing in Eq.~\eqref{eq:Oepsilon1}.
The inverse matrix of $\mC$ at $\by=\by^{(0)}$ is
\begin{equation}
  \begin{split}
    & [\mC(\by^{(0)})]^{-1} = \dfrac{1}{M_{2}^{2}-M_{1}^{2}} \\
    & \times
    \begin{pmatrix}
      M_{2} & -M_{1}\cos\phi^{(0)} & \frac{1}{l_{\ast}} M_{1}\sin\phi^{(0)} \\
      -M_{1}\cos\phi^{(0)} & M_{2} & \frac{1}{l_{\ast}} M_{1}\sin\phi^{(0)} \\
      \frac{1}{l_{\ast}} M_{1}\sin\phi^{(0)} & \frac{1}{l_{\ast}} M_{1}\sin\phi^{(0)}
      & \frac{2}{l_{\ast}^{2}}(M_{2}+M_{1}\cos\phi^{(0)}) \\
    \end{pmatrix}.
  \end{split}
\end{equation}
The matrix $\mX(\by^{(0)})=[\mC(\by^{(0)})]^{-1}\mK$ is hence
\begin{equation}
  \mX(\by^{(0)}) = \dfrac{k}{M_{2}^{2}-M_{1}^{2}}
  \begin{pmatrix}
    M_{2} & -M_{1}\cos\phi^{(0)} & 0 \\
    -M_{1}\cos\phi^{(0)} & M_{2} & 0 \\
    \frac{1}{l_{\ast}} M_{1}\sin\phi^{(0)} & \frac{1}{l_{\ast}} M_{1}\sin\phi^{(0)} & 0 \\
  \end{pmatrix}.
\end{equation}

\section{Validity of the hypothesis}
\label{sec:hypothesis-check}

Under the equal mass condition $m_{2}=m$,
we examine validity of the hypothesis ({\it H}) expressed in Eq.~\eqref{eq:hypothesis}.
We introduce the approximations of
\begin{equation}
  \dfracp{l_{j}^{(1)}}{t_{0}} \to \dfracd{l_{j}}{t},
  \quad
  \phi^{(0)} \to \phi,
\end{equation}
and the amplitudes of normal modes are expressed as
\begin{equation}
  \left\{
    \begin{split}
      w_{1}^{2}
      & = \dfrac{1}{2} \left[
        \left( l_{1} + l_{2} - 2l_{\ast} \right)^{2}
        + \dfrac{1}{\lambda_{1}}
        \left( \dot{l}_{1} + \dot{l}_{2} \right)^{2} \right],
      \\
      w_{2}^{2}
      & = \dfrac{1}{2} \left[
        \left( l_{1} - l_{2} \right)^{2}
        + \dfrac{1}{\lambda_{2}}
        \left( \dot{l}_{1} - \dot{l}_{2} \right)^{2} \right],
      \\
    \end{split}
  \right.
\end{equation}
where the eigenvalues $\lambda_{1}$ and $\lambda_{2}$
are defined in Eq.~\eqref{eq:X-eigenvalues}.
We compute the normal mode energy ratio defined by
\begin{equation} 
  R = \dfrac{E_{1}}{E_{1}+E_{2}},
  \quad
  E_{j} = \dfrac{k}{2}w_{j}^{2}~(j=1,2).
\end{equation}
The hypothesis is valid if $R$ is constant in time.

\begin{figure}
  \centering
  \includegraphics[width=8cm]{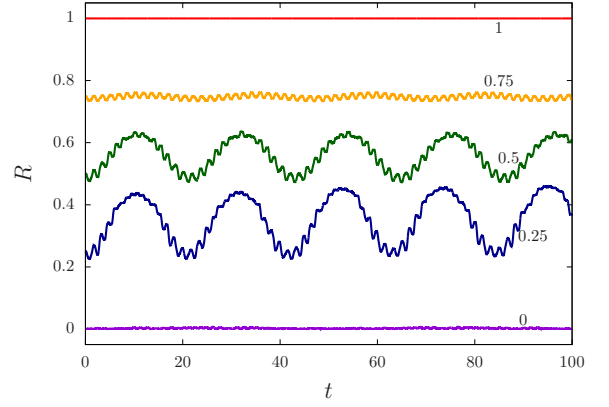}
  \caption{Temporal evolution of the normal mode energy ratio $R$.
    $\nu_{1}=1.0$ (red), $0.75$ (orange), $0.5$ (green), $0.25$ (blue),
    and $0$ (magenta) from top to bottom.
    The amplitude of the normal modes is $w=1.5$.
  }
  \label{fig:ModeEnergyRatio}
\end{figure}

We use the initial condition of Eq.~\eqref{eq:initial-condition}
with $\phi_{\ast}=\pi/2$,
and the amplitude of the normal modes is $w=1.5$.
Temporal evolution of $R$ is exhibited in Fig.~\ref{fig:ModeEnergyRatio}
for $\nu_{1}=1,~0.75,~0.5,~0.25$, and $0$ with $\nu_{2}=1-\nu_{1}$.
The hypothesis ({\it H}) is valid around $\nu_{1}=1$ and $0$ in particular.

\section{Analysis of the effective potential $\Veff$}
\label{sec:Zmu-analysis}

Let us study the critical points of the effective potential $\Veff$.
A critical point is defined as the point at which $\Veff'=0$.
The derivative of $\Veff$ is
\begin{equation}
  \Veff'(\phi) = \Meff(\phi) G_{\nu}(\phi).
\end{equation}
Thus, we have
\begin{equation}
  \phi \text{ is a critical point}
  \quad\Longleftrightarrow\quad
  G_{\nu}(\phi) = 0
\end{equation}
since the effective mass $\Meff$ is always positive as found in Eq.~\eqref{eq:Meff}.
The effective potential at a critical point takes a local minimum or a local maximum
depending on the sign of the second derivative.
At a critical point, we have $G_{\nu}=0$ and the second derivative of $\Veff$ is
\begin{equation}
  \phi \text{ is a critical point}
  \quad\Longrightarrow\quad
  \Veff''(\phi) = \Meff(\phi) G_{\nu}'(\phi).
\end{equation}
Again from $\Meff>0$, the sign of $\Veff''$ is determined by the sign of $G_{\nu}'$.
Keeping in mind the above discussions,
we study the critical points of the effective potential
for absence and appearance of the bending potential.

\subsection{Absence of the bending potential}

The function $G_{\nu}$ is proportional to $T_{\nu}$ for $\Vbend\equiv 0$,
and the sinusoidal function in $T_{\nu}$ gives the two critical points of
$\phi=0$ and $\pi$. In addition, there are the other two possible critical points
$\phi=\pm\phi_{\sharp}$ which solve the equation
\begin{equation}
  \dfrac{\nu_{1}}{M_{2}-M_{1}\cos\phi} - \dfrac{\nu_{2}}{M_{2}+M_{1}\cos\phi} = 0
\end{equation}
with $\nu_{2}=1-\nu_{1}$ and exist in the interval
\begin{equation}
  \label{eq:nu1-interval}
  \dfrac{M_{2}-M_{1}}{2M_{2}} < \nu_{1} < \dfrac{M_{2}+M_{1}}{2M_{2}}.
\end{equation}

The derivatives of $G_{\nu}$ at $\phi=0$ and $\pi$ are respectively
\begin{equation}
  \label{eq:Gnud0}
  G_{\nu}'(0) = \dfrac{E^{(2)}}{l_{\ast}^{2}}
  \dfrac{M_{1}}{M_{2}-M_{1}} \dfrac{(2\nu_{1}-1)M_{2}+M_{1}}{M_{2}^{2}-M_{1}^{2}}
\end{equation}
and
\begin{equation}
  \label{eq:Gnudpi}
  G_{\nu}'(\pi) = \dfrac{E^{(2)}}{l_{\ast}^{2}}
  \dfrac{M_{1}}{M_{2}+M_{1}} \dfrac{(1-2\nu_{1})M_{2}+M_{1}}{M_{2}^{2}-M_{1}^{2}},
\end{equation}
where we used the relation $\nu_{2}=1-\nu_{1}$.
The point $\phi=0$ is hence a local minimum point ($\Veff''(0)>0$) if and only if
\begin{equation}
  \nu_{1} > \dfrac{M_{2}-M_{1}}{2M_{2}},
\end{equation}
and the point $\phi=\pi$ is a local minimum point ($\Veff''(\pi)>0$) if and only if
\begin{equation}
  \nu_{1} < \dfrac{M_{2}+M_{1}}{2M_{2}}.
\end{equation}
The two points $\phi=0$ and $\phi=\pi$ are the local minimum points
in the interval of Eq.~\eqref{eq:nu1-interval}.
The periodicity of the effective potential $\Veff(\phi)$
requires the same numbers of local minimum points ($\Veff''>0$)
and local maximum points ($\Veff''<0$),
and hence the critical points $\phi=\pm\phi_{\sharp}$ are the local maximum points.

\subsection{Appearance of the bending potential}

We write the function $G_{\nu}(\phi)$ as
\begin{equation}
  G_{\nu}(\phi) = \dfrac{\sin\phi}{l_{\ast}^{2}(M_{2}-M_{1}\cos\phi)} g_{\nu_{1}}(\phi)
\end{equation}
where
\begin{equation}
  \begin{split}
    g_{\nu_{1}}(\phi) =
    & -8 \cos\phi + (E^{(2)}-\cos 2\phi - 1) \\
    & \times \left[
    \dfrac{\nu_{1}}{M_{2}-M_{1}\cos\phi}
    - \dfrac{\nu_{2}}{M_{2}+M_{1}\cos\phi}
  \right].
  \end{split}
\end{equation}
The effective potential has the critical points
at $\phi=0,\pi$, and $\phi_{\sharp}$ satisfying $g(\phi_{\sharp})=0$.
We separately discuss the second derivative
\begin{equation}
  \label{eq:ddVeff}
  \begin{split}
    \Veff''(\phi)
    & = g_{\nu_{1}}(\phi)
    \dfracp{}{\phi} \left[ \dfrac{\sin\phi}{l_{\ast}^{2}(M_{2}-M_{1}\cos\phi)} \right] \\
    & + \dfrac{\sin\phi}{l_{\ast}^{2}(M_{2}-M_{1}\cos\phi)} g_{\nu_{1}}'(\phi)
  \end{split}
\end{equation}
at a critical point.

\subsubsection{The critical points $\phi=0$ and $\pi$}
The second term of the right-hand side of Eq.~\eqref{eq:ddVeff} is zero,
and
\begin{equation}
  \dfracp{}{\phi} \left[ \dfrac{\sin\phi}{l_{\ast}^{2}(M_{2}-M_{1}\cos\phi)} \right]
  = \dfrac{\cos\phi}{l_{\ast}^{2}(M_{2}-M_{1}\cos\phi)}.
\end{equation}
This factor is positive (negative) at $\phi=0$ ($\pi$),
and the sign of $\Veff''$ is determined by $g_{\nu_{1}}(\phi)$.
We have
\begin{equation}
  g_{\nu_{1}}(0) = -8 + (E^{(2)}-2) \dfrac{M_{1}M_{2}}{M_{2}^{2}-M_{1}^{2}}
  \left( 2 \nu_{1} - 1 + \dfrac{M_{1}}{M_{2}} \right)
\end{equation}
and
\begin{equation}
  g_{\nu_{1}}(\pi) = 8 + (E^{(2)}-2) \dfrac{M_{1}M_{2}}{M_{2}^{2}-M_{1}^{2}}
  \left( 2 \nu_{1} - 1 - \dfrac{M_{1}}{M_{2}} \right).
\end{equation}
We separately discuss for $0\leq E^{(2)}\leq 2$ and $E^{(2)}>2$,
where the boundary $E^{(2)}=2$ comes from the bending potential
at $\phi=0$ and $\pi$: $\Vbend^{(2)}(0)=\Vbend^{(2)}(\phi)=2$.

If $0\leq E^{(2)}\leq 2$, the maximum value of $g_{\nu_{1}}(0)$ is realized at $\nu_{1}=0$,
which gives
\begin{equation}
  g_{0}(0) = -8 + (2-E^{(2)}) \dfrac{1-1/(M_{2}/M_{1})}{M_{2}/M_{1}-1/(M_{2}/M_{1})}
  < -7
\end{equation}
for $0\leq E^{(2)}\leq 2$. Note $M_{2}/M_{1}>1$.
The point $\phi=0$ is hence a local maximum point for $0\leq E^{(2)}\leq 2$.
A similar discussion states that the point $\phi=\pi$ is a local maximum point
for $0\leq E^{(2)}\leq 2$.

For $E^{(2)}>2$, the effective potential takes a local minimum at $\phi=0$ if
\begin{equation}
  \label{eq:0-stable}
  \nu_{1} > \dfrac{1}{2} \left( 1 - \dfrac{1}{M_{2}/M_{1}} \right)
  + \dfrac{4}{E^{(2)}-2} \left( \dfrac{M_{2}}{M_{1}} - \dfrac{1}{M_{2}/M_{1}} \right),
\end{equation}
and takes a local minimum at $\phi=\pi$ if
\begin{equation}
  \label{eq:pi-stable}
  \nu_{1} < \dfrac{1}{2} \left( 1 + \dfrac{1}{M_{2}/M_{1}} \right)
  - \dfrac{4}{E^{(2)}-2} \left( \dfrac{M_{2}}{M_{1}} - \dfrac{1}{M_{2}/M_{1}} \right).
\end{equation}
By changing the inequalities into the equality,
Eq.~\eqref{eq:0-stable} gives the red dotted-line
and Eq.~\eqref{eq:pi-stable} gives the blue dashed-line
in Figs.~\ref{fig:PhaseDiagramEqualMass} and \ref{fig:PhaseDiagram}.
We remark that Eq.~\eqref{eq:pi-stable} is equivalent with
\begin{equation}
  \nu_{2} > \dfrac{1}{2} \left( 1 - \dfrac{1}{M_{2}/M_{1}} \right)
  + \dfrac{4}{E^{(2)}-2} \left( \dfrac{M_{2}}{M_{1}} - \dfrac{1}{M_{2}/M_{1}} \right),
\end{equation}
whose right-hand side is identical with that of Eq.~\eqref{eq:0-stable}.

\subsubsection{The critical point $\phi=\phi_{\sharp}$}
We have $g(\phi_{\sharp})=0$, and
\begin{equation}
  \Veff''(\phi_{\sharp})
  = \dfrac{\sin\phi_{\sharp}}{l_{\ast}^{2}(M_{2}-M_{1}\cos\phi_{\sharp})}
  g'(\phi_{\sharp}).
\end{equation}
The effective potential takes a local minimum (maximum)
if $\Veff''(\phi_{\sharp})>0$ ($\Veff''(\phi_{\sharp})<0$)
at the critical point $\phi_{\sharp}$.

\section{Explanation on movies}
\label{sec:movie}

We provide movies for the initial conformations shown in Fig.~\ref{fig:MovieExplanation}
in Supplemental Material \cite{SM}.
The bending potential $\Vbend^{(2)}(\phi)$ of Eq.~\eqref{eq:Vbend-example} is in use.
The excited spring mode is $(\nu_{1},\nu_{2})=(1,0)$,
and the amplitude $w$ is $w=0.5$ for Figs.~\ref{fig:MovieExplanation} (a) and (d),
$w=1$ for (b) and (e), and $w=1.5$ for (c) and (f).
The other initial conditions are described in Sec.~\ref{sec:initial-condition},
and the system parameter values ($m_{i},k$, and $l_{\ast}$)
are given in Sec.~\ref{sec:setting}.
Dynamics of the system corresponding to the panels
from Figs.~\ref{fig:MovieExplanation}(a) to (f)
is demonstrated in from MovieA to MovieF, respectively.
We stress that temporal evolution is well understood by the effective potential $\Veff$,
while the bending potential [see Fig.~\ref{fig:Veff}(a)] does not explain it.

\begin{figure}[h]
  \centering
  \includegraphics[width=8cm]{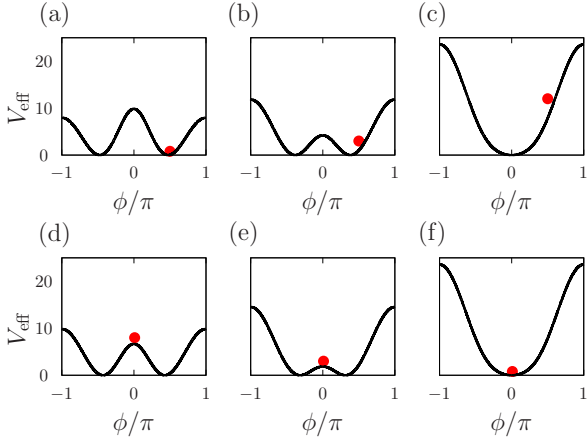}
  \caption{Initial conformations (red points) and effective potentials (black lines).
    The reference conformations are $\phi_{\ast}=\pi/2$ in the panels (a), (b), and (c),
    and $\phi_{\ast}=0.01$ in (d), (e), and (f),
    where the initial conformation $\phi_{0}$ is close to $\phi_{\ast}$.
    See Figs.~\ref{fig:phi_phiinipiover2}(d), (e), and (f)
    for temporal evolution of $\phi(t)$ corresponding to the panels (a), (b), and (c),
    respectively.
  }
  \label{fig:MovieExplanation}
\end{figure}

\section{From a general potential to the bending potential}
\label{sec:GeneralPotential}

The bending potential $\Vbend(\phi)$ represents
the interaction between the two beads of the ends,
because the interction between an end and the center beads
results in the spring potential.
We show that the bending potential energy function of $\phi$
is derived from any potential $V_{\rm G}$ which is a function of
the distance $r=\norm{\br_{3}-\br_{1}}$,
although it is assumed to be a function of only $\phi$ in the main text.
Here $r$ is represented by using $l_{1}, l_{2}$, and $\phi$ as
\begin{equation}
  \label{eq:r-l-phi}
  r = \norm{(\br_{3}-\br_{2})+(\br_{2}-\br_{1})}
  = \sqrt{l_{1}^{2} + l_{2}^{2} + 2l_{1}l_{2} \cos\phi}.
\end{equation}

Substituting Eq.~\eqref{eq:l-phi-expansion} into Eq.~\eqref{eq:r-l-phi},
we have $r = r^{(0)} + O(\epsilon)$ and
\begin{equation}
  r^{(0)} = l_{\ast} \sqrt{2(1+\cos\phi^{(0)})}.
\end{equation}
As shown in Appendix \ref{sec:Vbend-Oepsilon2},
the potential $V_{\rm G}$ is of $O(\epsilon^{2})$
under the assumptions {\it (A1)} and {\it (A2)},
and we expand it as
\begin{equation}
  V_{\rm G}(r) = \epsilon^{2} V_{\rm G}^{(2)}(r) + O(\epsilon^{3})
  = \epsilon^{2} V_{\rm G}^{(2)}(r^{(0)}) + O(\epsilon^{3}).
\end{equation}
Therefore, the second order bending potential $\Vbend^{(2)}$
is derived as a function of only $\phi^{(0)}$ as
\begin{equation}
  \Vbend^{(2)}(\phi^{(0)}) = V_{\rm G}^{(2)} \left( l_{\ast} \sqrt{2(1+\cos\phi^{(0)})} \right).
\end{equation}
The effective potential $\Veff$ is obtained from Eq.~\eqref{eq:Veff}
by substituting the above bending potential $\Vbend^{(2)}$ into Eq.~\eqref{eq:Gnu}.

\end{document}